\documentclass[11pt,twoside]{article}
\usepackage{asp2010}

\resetcounters

\begin{document}

\title{Computing and Using Metrics in the ADS}
\author{Edwin A. Henneken,$^1$ Alberto Accomazzi,$^1$, Michael J. Kurtz$^1$, Carolyn S. Grant$^1$, Donna Thompson$^1$, Jay Luker$^1$, Roman Chyla$^1$, Alexandra Holachek$^1$ and Stephen S. Murray$^2$.
\affil{$^1$Smithsonian Astrophysical Observatory, 60 Garden Street, Cambridge MA 02138, USA},
\affil{$^2$Johns Hopkins University, Physics and Astronomy Department, 3400 N. Charles Street, Baltimore, MD 21218, USA,}
}

\begin{abstract}
Finding measures for research impact, be it for individuals, institutions, instruments or projects, has gained a lot of popularity. More papers than ever are being written on new impact measures, and problems with existing measures are being pointed out on a regular basis. Funding agencies require impact statistics in their reports, job candidates incorporate them in their resumes, and publication metrics have even been used in at least one recent court case. To support this need for research impact indicators, the SAO/NASA Astrophysics Data System (ADS) has developed a service which provides a broad overview of various impact measures. In this presentation we discuss how the ADS can be used to quench the thirst for impact measures. We will also discuss a couple of the lesser known indicators in the metrics overview and the main issues to be aware of when compiling publication-based metrics in the ADS, namely author name ambiguity and citation incompleteness.
\end{abstract}

\section{Introduction}
The study of quantitative aspects of (scholarly) publishing and publications (``bibliometrics'' and ``scientometrics'') has been around for decades. But it is since recent times that this interest is more than purely academic. The fact that there has been a sharp increase in publications on the various incarnations of ``informetrics'' in the last decade is a reflection of this trend (see e.g. \citet{Larivière2012}). In this period the quest for being able to quantify ``research impact'' has increased as well. Since citations form the currency of scholarly publishing, it is not surprising that citation-based indicators have emerged as the building blocks for a wide variety of ``impact factors'' or ``metrics''. Usage, for example in the form of download statistics, is another signal available for constructing indicators, but it is inherently more noisy than citations. However, usage-based indicators have been shown to be relevant additions to citation-based indicators (see \citet{KurtzBollen2010}, \citet{Kurtz2005}). The frequent requests for citation statistics made it clear that there was a need for a service that would generate an overview of statistics and indicators, based on a set of publications. This service is available in ADS 2.0 (http://adslabs.org) and provides a set of ``canned'' statistics. ADS 2.0 also comes with a powerful new query syntax, which makes it singularly useful for more custom bibliometric analyses and impact studies.

\section{Computing Metrics}
\subsection{Custom metrics}
Some measures of impact cannot be generated by simply selecting a set of records and generating the ``canned'' metrics overview. The powerful query syntax makes it relatively easy to explore complex questions and generate statistics based on the results. Questions like ``what is the percentage of the refereed astronomy literature, published through the core astronomy journals, in a given year?'' and ``how has the role of a given journal changed over time for a given field?'' are examples of such questions. Once you have identified what everything means in the questions, the next step is to translate them into the ADS 2.0 query language (http://labs.adsabs.harvard.edu/adsabs/page/help/search). To expore the first question, you fire off the following query:
\begin{quote}
database:"astronomy" year:2013 property:"refereed"\footnote{NOTE: By default search terms will be combined using AND as the default boolean operator, but this can be changed by explicitly specifying OR beween them (which was the default in ADS Classic)}
\end{quote}
which returns all refereed publications in the ADS astronomy database for a given year (2013 in this case). The results page then essentially answers your initial question. It provides you with the total number of refereed astronomy publications for that year, and the ``Publications'' facet tells you how many publications among these results were published in which journals. If you want to further explore this question with the follow-up question: ``of all refereed citations to publications in the refereed astronomy literature, for a given year, what is the percentage of citations to the core astronomy journals?''. An essential ingredient to answer this question is to find all citations to, for example, articles in the Monthly Notices of the R.A.S., which would be retrieved as follows:
\begin{quote}
citations(bibstem:"MNRAS" year:2013) property:"refereed"
\end{quote}
where the 'bibstem' modifier is used to specify the publication (using the ADS journal abbreviation).
It may be interesting for publishers to know if there are any trends in their journals with respect to subject matter. For authors, these are interesting questions too, because it can help them decide where to submit a paper. For example, you could wonder if there is a trend in the Astrophysical Journal (main section) with respect to papers about ``weak lensing''. A question like this really consists of, at least, two components. First you would want to know if there is a trend in the percentage of articles (both within the journal itself, and within the field). Second, it would be interesting to know how articles on this subject are being cited. To address the first part of the question, you would use a query like
\begin{quote}
"weak lensing" bibstem:"ApJ" -page:"L*" year:2013
\end{quote}
where the minus sign, prepended to the 'page' modifier, is used to indicate that the results returned should \textbf{not} have this attribute. By default, the system searches both metadata \textbf{and} the full article text (if available). If you wish the above query to just search title or abstract, you would use the query
\begin{quote}
(title:"weak lensing" OR abs:"weak lensing") bibstem:"ApJ" -page:"L*" year:2013
\end{quote}
To see how articles on ``weak lensing'' (in a given year) cite articles in the Astrophysical Journal (main section), you would use the query
\begin{quote}
references("weak lensing" year:2013) bibstem:"ApJ" -page:"L*"
\end{quote}
In the section on using the ADS metrics, we will describe how these numbers can be generated in a programmatic way, so that you do not have to copy and paste results from your browser.
\subsection{Canned metrics}
The ``canned'' metrics service generates an overview of citation and usage statistics, together with a number of derived indicators (like the Tori and Read10 indices). This service would be useful to answer questions like ``what are the publication statistics for Adam G. Riess, for first-authored, refereed papers, published in the time period 2000-2013?''. To generate the statistics for this particular question, you run the query
\begin{quote}
author:"\string^Riess, Adam G." year:2000-2013 property:"refereed"
\end{quote}
to generate all the records that match, and then you generate the metrics overview by selecting ``Metrics'' in the ``Analyze'' menu. If you do not select any records a dialog box will appear and you may select all of your results (to a maximum of 3000) or adjust the number of papers to be analyzed by using the slide bar. The overview that will be generated next, consists of a number of obvious statistics, a number of derived indicators, a set of histograms and a plot with time series for a number of indicators. Two derived indicators may be somewhat unfamiliar, so we will briefly illustrate their meaning. The $tori$ (``total research impact'') index (see~\citet{Pepe2012}) quantifies, for an individual, the total amount of scholarly work that others have devoted to his/her work, measured in the volume of research papers. 
\begin{equation}
\label{formula:ror}
tori = \displaystyle\sum_{i}\sum_{c\in C_i} \frac{1}{a_{i} \cdot n_c}
\end{equation}
where $i$ runs over the set of articles, $C_i$ is the set of citations (excluding self-citations) for article $i$, $a_i$ is the number of authors in article $i$ and $n_c$ is the number of references in citation $c$. In other words, if a researcher writes many single-authored papers which get cited a lot by papers with small bibliographies, this researcher's tori index is very likely to be relatively high. 
The other indicator we would like to highlight is the $Read\mathit{10}$ index. This index is the current readership rate for all the papers published by an author in the most recent ten years, normalized by number of authors in each paper. This means that this indicator, for the first 10 years of a researcher's career, equals the regular, author-normalized readership rate. Once a researcher has stopped publishing, his/her $Read\mathit{10}$ will drop to zero after 10 years. This indicator is a measure for an individual's current activity and is therefore a very useful addition to citation-based indicators.
\section{Using Metrics}
Visually exploring is insightful by itself, but eventually you will want to be able to retrieve data in some automated fashion, rather than copy-and-pasting it from your browser! This is very simple in the case of our ``canned'' metrics report. This overview comes with two buttons, allowing you to dowload the numbers as an Excel file or in the form of a PDF report. If you feel comfortable with writing Python scripts, the new ADS API will enable you to do queries and retrieve statistics. All necessary information for the ADS API can be found here:
\begin{quote}
https://github.com/adsabs/adsabs-dev-api
\end{quote}
To obtain access to the ADS Developer API you must do two things. First login to ADS 2.0 (upper right corner). We encourage you to use your existing "Classic ADS" login credentials if you already have an ADS account. If not, you can create a new account via ADS 2.0. Next, you apply for an API developer token. The form to apply for this token can be reached on the above URL. All API requests must include your developer token.
\section{Discussion}
The use of any metric or impact factor is controversial for many reasons. The following quote from Albert Einstein sums up a part of the problem: ``Not everything that can be counted counts, and not everything that counts can be counted''. But even if this were not a problem, there are still numerous factors that make the construction of any indicator complicated. For one, there is no such thing as a complete bibliographic database. The scope and coverage of any database that can be used for creating indicators is not exhaustive. For example, if you are looking for publication statistics for the Nobel laureate Charles K. Kao, and you would use a database that does not cover IEEE publications, you would vastly undercount his citations. When publications are retrieved based on author names, there will always be the problem of ambiguity or name changes, which hopefully will get remedied by the introduction of an author ID system (like, for example, ORCID). In addition, citations are not created equal. In the realm of citations, no distinction is made between praise and criticism. Also, the use of citations to contruct indicators penalizes those researchers who have long-term projects that result in few publications, even though such long-term studies might be very important. There is also the problem of discipline-dependent citation pratices. Some citation-based indicators can be easily contaminated by practices like abundant self-citation and the creation of ``citation clubs''. There are notorious examples of authors that cite themselves over 200 times within one SPIE conference proceeding. Any ``metrics'' report should have Caveat Emptor imprinted on it.

\acknowledgements The ADS is funded by NASA grant NNX12AG54G.

\end{document}